\documentclass[superscriptaddress,twocolumn,10pt,floatfix]{revtex4-1}

\usepackage[english]{babel}
\usepackage{amsmath}  % needed for \tfrac, \bmatrix, etc.
\usepackage{amsfonts} % needed for bold Greek, Fraktur, and blackboard bold
\usepackage{graphicx} % needed for figures
\usepackage{hyperref}
\graphicspath{ {./} }
\usepackage{array}
\usepackage[table]{xcolor}
\usepackage{ulem}
\usepackage{xcolor}

\begin{document}

% Be sure to use the \title, \author, \affiliation, and \abstract macros
% to format your title page.  Don't use lower-level macros to  manually
% adjust the fonts and centering old title:Parity readout of spin qubits in silicon quantum dots
\title{Pauli blockade in silicon quantum dots with spin-orbit control}
\author{Amanda E. Seedhouse}\thanks{Corresponding Authors}\affiliation{School of Electrical Engineering and Telecommunications, The University of New South Wales, Sydney, NSW 2052, Australia}
\author{Tuomo Tanttu}\thanks{Corresponding Authors}\affiliation{School of Electrical Engineering and Telecommunications, The University of New South Wales, Sydney, NSW 2052, Australia}
\author{Ross C. C. Leon}\affiliation{School of Electrical Engineering and Telecommunications, The University of New South Wales, Sydney, NSW 2052, Australia}

\author{Ruichen Zhao} \thanks{Current address: National Institute of Standards and Technology, 325 Broadway, Boulder, CO, USA, 80305.}\affiliation{School of Electrical Engineering and Telecommunications, The University of New South Wales, Sydney, NSW 2052, Australia}

\author{Kuan Yen Tan}\thanks{Current address: IQM Finland Oy, Keilaranta 19, 02150 Espoo, Finland.}\affiliation{QCD Labs, QTF Centre of Excellence, Department of Applied Physics, Aalto University, 00076 AALTO, Finland}

\author{Bas Hensen}\thanks{Current address: TU delft, Netherlands}\affiliation{School of Electrical Engineering and Telecommunications, The University of New South Wales, Sydney, NSW 2052, Australia}
\author{Fay E. Hudson} \affiliation{School of Electrical Engineering and Telecommunications, The University of New South Wales, Sydney, NSW 2052, Australia}
\author{Kohei M. Itoh}\affiliation{School of Fundamental Science and Technology, Keio University,3-14-1 Hiyoshi, Kohoku-ku, Yokohama 223-8522, Japan}
\author{Jun Yoneda}\affiliation{School of Electrical Engineering and Telecommunications, The University of New South Wales, Sydney, NSW 2052, Australia}
\author{Chih Hwan Yang}\affiliation{School of Electrical Engineering and Telecommunications, The University of New South Wales, Sydney, NSW 2052, Australia}
\author{Andrea Morello}\affiliation{School of Electrical Engineering and Telecommunications, The University of New South Wales, Sydney, NSW 2052, Australia}
\author{Arne Laucht}\affiliation{School of Electrical Engineering and Telecommunications, The University of New South Wales, Sydney, NSW 2052, Australia}
\author{Susan N. Coppersmith}\affiliation{School of Physics, University of New South Wales, Sydney, NSW 2052, Australia}
\author{Andre Saraiva}\thanks{Corresponding Authors}\affiliation{School of Electrical Engineering and Telecommunications, The University of New South Wales, Sydney, NSW 2052, Australia}
\author{Andrew S. Dzurak}\thanks{Corresponding Authors}\affiliation{School of Electrical Engineering and Telecommunications, The University of New South Wales, Sydney, NSW 2052, Australia}
\date{\today}

\begin{abstract}
Quantum computation relies on accurate measurements of qubits not only for reading the output of the calculation, but also to perform error correction. Most proposed scalable silicon architectures utilize Pauli blockade of triplet states for spin-to-charge conversion. In recent experiments there have been instances when instead of conventional triplet blockade readout, Pauli blockade is sustained only between parallel spin configurations, with $|T_0\rangle$ relaxing quickly to the singlet state and leaving $|T_+\rangle$ and $|T_-\rangle$ states blockaded -- which we call \textit{parity readout}. Both types of blockade can be used for readout in quantum computing, but it is crucial to maximize the fidelity and understand in which regime the system operates. We devise and perform an experiment in which the crossover between parity and singlet-triplet readout can be identified by investigating the underlying physics of the $|T_0\rangle$ relaxation rate. This rate is tunable over four orders of magnitude by controlling the Zeeman energy difference between the dots induced by spin-orbit coupling, which in turn depends on the direction of the applied magnetic field. We suggest a theoretical model incorporating charge noise and relaxation effects that explains quantitatively our results. Investigating the model both analytically and numerically, we identify strategies to obtain on demand either singlet-triplet or parity readout consistently across large arrays of dots. We also discuss how parity readout can be used to perform full two-qubit state tomography and its impact on quantum error detection schemes in large-scale silicon quantum computers.
\end{abstract}


\maketitle

\section{Introduction}
The recent demonstration of large-scale quantum computation~\cite{arute_quantum_2019} has opened the door to the exploration of near-term applications of noisy, intermediate-scale devices. This, however, does not change the long-term vision wherein quantum error correction is essential to achieve the full advantages of quantum computing~\cite{knill_theory_1997}. Theoretical estimates predict a large overhead in terms of the number of required physical qubits for this task~\cite{steane_overhead_2003}. As quantum computing technologies progress from demonstrations to industrial platforms, silicon-based architectures become increasingly competitive due to the possibility of mass production of few-nanometer-sized qubit systems~\cite{cai_silicon_2019, veldhorst_silicon_2017,helmer_cavity_2009, maurand2016cmos}. Such qubits have high control fidelity at the one-~\cite{yoneda_quantum-dot_2018, yang_silicon_qubit_2019,kawakami_gate_2016,muhonen_quantifying_2015,takeda_fault-tolerant_2016,kim_high-fidelity_2015} and two-qubit~\cite{huang_fidelity_2019, xue_benchmarking_2019, Zajac_resonantly_2018} levels.

Error correcting codes require a highly connected network of qubits~\cite{wang2011surface}, setting topological constraints on large scale designs. Pauli spin blockade can be employed to readout the state of silicon qubits~\cite{cai_silicon_2019, veldhorst_silicon_2017, jones_logical_2018}, which removes the need for reservoirs near the dots, and loosens design constrains. Recent experiments have validated the operation of single spin qubits together with Pauli blockade readout in silicon-metal-oxide-semiconductor (SiMOS) quantum dots~\cite{fogarty_integrated_2018,zhao_single-spin_2019,yang_silicon_2019}. In some experiments~\cite{yang_silicon_2019}, this singlet-triplet readout has not been complete. The non-polarized triplet $T_0$ state in these experiments decays faster than the measurement bandwidth limits, leading to a \textit{parity readout} \cite{Engel_fermionic_2005} where the blockade is selective to whether the total parity of the spins is odd or even. This result is not consistent for all silicon devices of the same type, since some silicon samples preserve singlet-triplet blockade~\cite{fogarty_integrated_2018,zhao_single-spin_2019}, raising the question: what conditions give rise to parity readout? This question becomes especially important in large arrays of dots, since some readout pairs might end up being in parity and some in singlet-triplet readout mode. Additionally, qubit encoded based on pairs of spins in which the logical 0 and 1 states are represented by singlets and triplets respectively, cannot be read out using parity readout~\cite{taylor_relaxation_2007}.

Here, we explore a device that operates at the interface of the parity and singlet-triplet readout, allowing us to study the origin of the $T_0$ blockade lifting. We can tune the $T_0$ blockade lifting rate by four orders of magnitude by controlling Zeeman energy difference, which is tuned in our system by varying the angle of the external magnetic field relative to the crystal lattice, since the Zeeman energy difference is dominated by spin-orbit interaction~\cite{tanttu_controlling_2019}. Incorporating this observation, we use perturbation theory to model the blockade time in the system both analytically~\cite{schrieffer_relation_1966} and computationally. We investigate both charge relaxation and dephasing as the mechanisms for $T_0$ blockade lifting. We find that both models fit qualitatively the experimental data, but the fitted charge dephasing time is more reasonable than the fitted relaxation time, suggesting a different type of underlying mechanism in silicon than in GaAs~\cite{taylor_relaxation_2007}. 

These results are an important tool to be able to understand how to tune the readout from parity to singlet-triplet and vice versa. Due to the statistical variability of Zeeman energies in an array of dots, gate control of tunnel rates between each pair of dots is necessary. The detuning at which readout is performed can also be used to control the readout regime, but the range of a workable detuning is limited by valley or orbital excitations. Moreover, the magnetic field angle can be set to point along (100), which removes the Dresselhauss contribution to the spin-orbit coupling~\cite{tanttu_controlling_2019}, consequently reducing the overall variability of g-factors between dots (which will then be dominated only by Rashba effect). The particular values of tunnel coupling required for each qubit pair are obtained within our theoretical analysis. We also explain how parity and singlet-triplet readout may be more appropriate for scaling up architectures compared to the latched Pauli spin blockade readout, or Elzerman readout which require routing reservoirs to nearby the dots. Finally we discuss how to perform full two-qubit state tomography as well as error detection~\cite{cai_silicon_2019, veldhorst_silicon_2017, jones_logical_2018}, utilizing parity readout. 

\begin{figure*}[t]
\includegraphics[width=16cm]{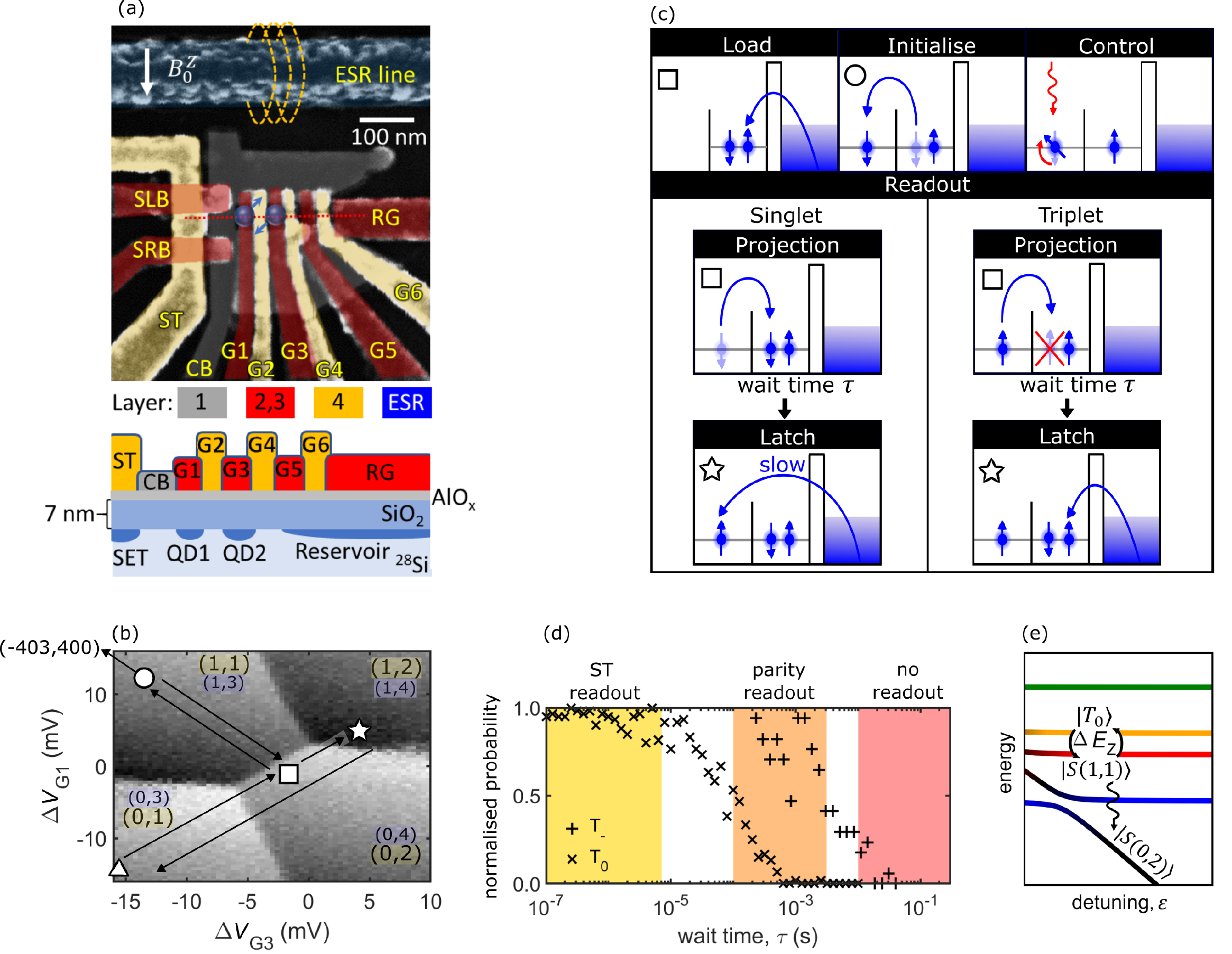}
\centering
\caption{Pauli spin blockade and latched readout. (a) False-colored scanning electron microscope (SEM) image of a device nominally identical to the one used in the experiments. The dashed line shows the location of the cross section of the device represented at the bottom of the image. (b) The charge stability diagram showing the electron occupancy in the device. The valence electrons in each dot are shown by the numbers highlighted in yellow; the actual electron occupancy is shown by the numbers highlighted in blue. The arrows show the different loading and unloading stages of the experiment and the symbols represent the different stages of the experiment. The circle represents the initialization point which is beyond the region of the stability diagram shown here (the arrow pointing out of the figure represents this). (c) A diagram representing the different stages of the experimental process and how the Pauli blockade allows for spin readout. The decay of $|T_0\rangle$ and $|T_-\rangle$ triplets is shown in (d) as a function of wait time $\tau$ at the blockade region near the (1,1)-(0,2) transition. We have normalized the probabilities of the decay of the states according to their initialization fidelity. (e) An energy diagram showing the Zeeman energy difference coupling the $|T_{0}(1,1)\rangle$ and $|S(1,1)\rangle$ states. The decay of $|S(1,1)\rangle$ into $|S(0,2)\rangle$ is shown by the arrow between the two states.} 
\label{fig:exp-setup}
\end{figure*}

%\section{Results}
\section{Experiments on rate of blockade lifting and g-factor difference}

Fig.~\ref{fig:exp-setup}(a) shows the scanning electron microscope image of a device nominally identical to the one used in the experiments, together with the schematic cross section. This device accumulates electrons in the isotopically purified silicon slab (800 ppm residual $^{29}$Si~\cite{itoh_isotope_2014}) and has been used in experiments reported in Ref.~\cite{zhao_single-spin_2019}. This device can be operated in singlet-triplet readout mode in contrast to device in Ref.~\cite{yang_silicon_2019} where due to the micromagnet only the parity readout was possible. Left and right dots are formed under gates G1 and G3 (made of palladium), respectively, and are laterally surrounded by the confinement barrier gate (CB). Figure \ref{fig:exp-setup}(b) depicts our operating regime near the (1,3)-(0,4) charge transition (the numbers in brackets represent the total amount of electrons in the left and right dots, respectively). The two lowest energy electrons in the right dot lie in the lower valley state, which is separated from the upper valley state by a large enough excitation energy to be easily discernible experimentally (typically larger than 0.1 meV). This results in a spin-0 closed shell that does not impact the spin dynamics of the two extra electrons. For simplicity, we ignore the two lowest energy electrons in the right dot and refer to the possible charge configurations as (1,1) or (0,2) later in the text.

The details of the measurement scheme are illustrated in Fig. \ref{fig:exp-setup}(c). The experiment starts in the (0,1) charge state. An electron is then loaded into the right dot, leading to a singlet ground state $|S(0,2)\rangle$. Then, the detuning between the dots is changed across the $(0,2)\rightarrow(1,1)$ transition, allowing one electron to tunnel into the left dot. The initial spin state depends on the ramp rate going from (0,2) to (1,1), which determines whether the energy anticrossings with $|T_-(1,1)\rangle$ or $|S(1,1)\rangle$ are swept adiabatically or not. Following this, the electrons are manipulated in different ways, either by electron spin resonance or detuning control. Finally, the resulting two-spin state would normally be measured by returning to the (0,2) charge configuration near the transition from (1,1) -- the readout configuration in the Pauli blockaded region~\cite{petta_pulsed-gate_2005}. This way, only the spin singlet would be able to tunnel back into the (0,2) state; therefore, the charge state of the double dot would reflect what the spin state was at the time of the measurement. This simplified picture, as we show next, only holds immediately after the dot is brought to the readout configuration.We initialize the $|T_0\rangle$ state with 97\% initialization fidelity by preparing a singlet state and consequently waiting for half the period of the spin-orbit induced singlet-triplet oscillations. The details of the experiment are explained in the Supplementary material in Ref.~\cite{zhao_single-spin_2019}. Alternatively, we also initialize $|T_-\rangle$ states by adiabatically preparing a $|\uparrow \downarrow\rangle$ state and pulsing a microwave in resonance with the transition in the left dot to $|\downarrow \downarrow\rangle$, which can be distinguished from the transition in the right dot  through spin-orbit interaction (see Ref.~\cite{tanttu_controlling_2019} for more details). This initialization has a limited visibility of around 30\% because the ESR antenna in the present device is defective.

To study the lifetime of the Pauli blockade of the spin triplet configurations, we use latched singlet-triplet readout~\cite{broome_high-fidelity_2017,harvey-collard_high-fidelity_2018,fogarty_integrated_2018, zhao_single-spin_2019}. This process is schematically illustrated in Fig.~\ref{fig:exp-setup}(c). After some wait time $\tau$ in the (0,2) readout configuration, we move to the latched readout region within the (1,2) configuration space.
Latching detects the difference in initial charge state, because of the slow loading of an electron to the left dot. The $|S(0,2)\rangle$ state will stay in the (0,2) configuration and the blockaded triplet states will allow the fast transition from the (1,1) into the (1,2) charge state. Now the charge state represents a mapping of the spin states at the moment when the latching pulse occurred. A slow charge sensing step will no longer compromise the conclusion regarding the spin state at the readout point. In summary, we end up with three possible readout schemes: latched Pauli spin blockade, singlet-triplet readout and parity readout. The former, latched readout, is not scalable but is used as a tool to investigate the other two readout mechanisms.

During the wait time in the Pauli blockade region, all triplet states will eventually decay to the ground state $|S(0,2)\rangle$. We observe that this rate is different for different triplet states, as shown in Fig.~\ref{fig:exp-setup}(d). The odd-parity triplet $|T_0\rangle$ decays in roughly 200 $\mu$s and the $|T_-\rangle$ decays in 5 ms. We mention, without showing here, that the $|T_+\rangle$ configuration also outlasts the $|T_0\rangle$ decay by orders of magnitude (see Ref.~\cite{yang_silicon_2019} for further evidence). If the charge sensor (a continuous current single electron transistor in our case) requires a data acquisition time between these two blockade lifting time scales, we do not distinguish between the two odd-parity spin configurations ($|S\rangle$ and $|T_0\rangle$), but we do distinguish between these and the even-parity configurations ($|T_+\rangle$ and $|T_-\rangle$). We call this a \textit{parity readout}.  This indicates that the physical mechanism for triplet blockade lifting is different between the even-parity triplets and $|T_0\rangle$. 

We describe our hypothesis. Even-parity triplet states require a spin flip process accompanied by the emission of a phonon in order to relax into the singlet ground state -- a process that is very slow in silicon~\cite{xiao_measurement_2010}. On the other hand, the triplet $|T_{0}(1,1)\rangle$ and the singlet $|S(1,1)\rangle$ belong to the same subspace spanned by the $\left|\uparrow\downarrow\rangle\right., \left|\downarrow\uparrow\rangle\right.$ states, which are approximately the system eigenstates in the (1,1) charge state, in the presence of a substantial difference in g-factor between the dots (see below). The g-factor difference induces a fast oscillation between $|T_{0}(1,1)\rangle$ and $|S(1,1)\rangle$. While at the readout position, the wavefunction component that evolves into the singlet $|S(1,1)\rangle$ state rapidly oscillates between that and the singlet $|S(0,2)\rangle$ state due to the fast interdot tunnel rate. Charge dephasing and phonon-mediated relaxation damp these oscillations, which results in a steady state in which both electrons eventually are found in the (0,2) state. This is made clearer in Fig. \ref{fig:exp-setup}(e) which shows a schematic energy diagram with the $|T_{0}(1,1)\rangle$ and $|S(1,1)\rangle$ oscillations, mediated through the Zeeman energy difference, and the decay into the $|S(0,2)\rangle$ state.

The $S-T_0$ mixing is caused by the difference in Zeeman energies $\Delta E_{\text{Z}}$ between the electrons in each dot. Since our device is fabricated on enriched $^{28}$Si, the Overhauser magnetic field is minimal and is not the main cause for $\Delta E_{\text{Z}}$ (as confirmed \textit{a posteriori}). Additionally, no magnetic materials were used for the metal stack and there is no Meissner effect since the gates are not superconducting. Therefore, $\Delta E_{\text{Z}}$ is mostly determined by the spatial variations of the effective Land\'e g factor due to the spin-orbit coupling induced by the interface. Under this approximation, we have $\Delta E_{\text{Z}} = \Delta g \mu_{\text{B}} B/h$, where the difference in g-factors between the two quantum dots is $\Delta g$, $\mu_{\text{B}}$ is the Bohr magneton constant, and $B$ is the magnitude of the magnetic field where $B$ = 600 mT for all experiments. These spin orbit effects include a large Dresselhaus component, which can be controlled by the direction of the external magnetic field with regard to the silicon lattice (the presence of an interface breaks the inversion symmetry of the lattice)~\cite{harvey2019spin, tanttu_controlling_2019}. We use this controllability to correlate the blockade rate and the difference in Zeeman energies.

\begin{figure*}[t]
\includegraphics[width=18cm]{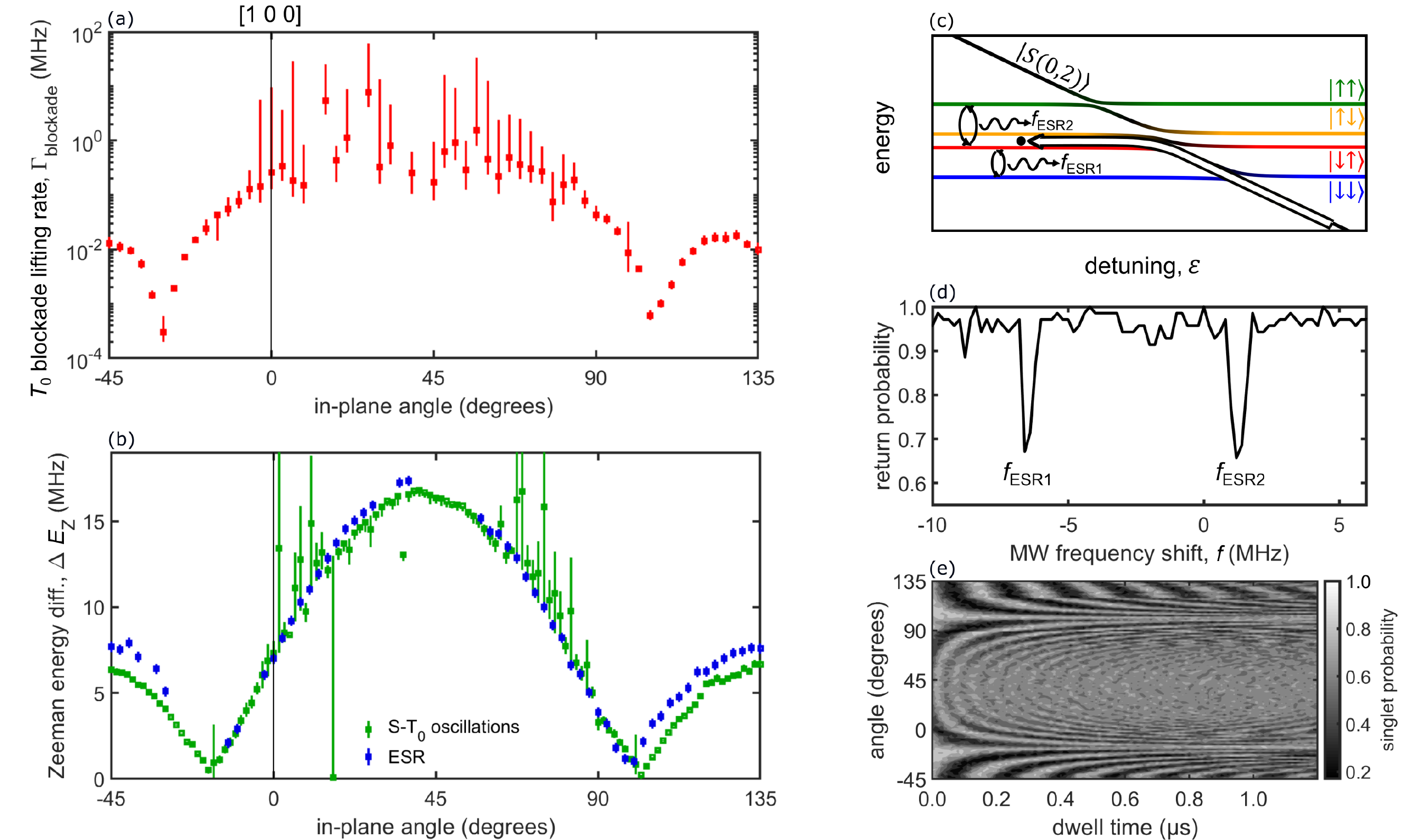}
\centering
\caption{Correlation between Zeeman energy difference and the rate of lifting the triplet blockade. (a) The rate of lifting the triplet blockade (red squares) extracted from decay plots similar to Fig.~\ref{fig:exp-setup}(d) and is plotted as a function of in-plane magnetic field angle for $B$ = 600 mT. (b) The difference in Zeeman energies plotted as a function of in-plane magnetic field angle, experimentally determined with two different methods, ESR (blue squares) and $S-T_{0}$ oscillations (green squares). (c) The energy diagram of a double dot system as a function of detuning, showing schematically the initialization and control strategy for each of the methods presented in (b) to find the Zeeman energies. The arrow shows the fast, diabatic plunge into (1,1) which initialises into $|S(1,1)\rangle$ and allows for observation of $S-T_{0}$ oscillations. The two frequencies $f_{\mathrm{ESR1}}$ and $f_{\mathrm{ESR2}}$ in (c) are shown as dips in (d), which are the individual resonance frequencies of the electron spins in each dot. (e) The $S-T_{0}$ oscillations as a function of dwell time and in-plane magnetic field angle.}
\label{fig:results}
\end{figure*}

Firstly, we repeat the analysis described in Fig.~\ref{fig:exp-setup}(d) for all magnetic field angles in the plane of the device. Figure~\ref{fig:results}(a) shows the extracted rate of blockade lifting, $\Gamma_{\text{blockade}}$, for $|T_0\rangle$ as red squares, which changes by 3 to 4 orders of magnitude. Next, the measurement of $\Delta E_{\text{Z}}$ is carried out using two methods, with results shown in Fig. \ref{fig:results}(b). One method is to use electron spin resonance (ESR) to rotate a single spin in one of the dots. This is accomplished by initialising $\left|\downarrow \uparrow \right\rangle$ by ramping adiabatically through the avoided crossing between $\left|\downarrow \uparrow \right\rangle$ and $\left|\uparrow \downarrow \right\rangle$ represented Fig. \ref{fig:results}(c). Then, a microwave pulse is applied using the ESR antenna which, at resonance frequencies $f_{\text{ESR1}}$ and  $f_{\text{ESR2}}$, may rotate $\left|\downarrow \uparrow \right\rangle$ to $\left|\uparrow \uparrow \right\rangle$ and $\left|\downarrow \uparrow \right\rangle$ to $\left|\downarrow \downarrow \right\rangle$, respectively (see Fig. \ref{fig:results}(c)). Returning to the positive detuning $\varepsilon$ configuration, the probability of a successful return to (0,2) will depend whether the ESR pulse was in resonance with one of the transitions. The frequencies that are resonant show a dip in the return probability plotted in Fig.~\ref{fig:results}(d). The difference in these two frequencies gives approximately $\Delta E_{\text{Z}}$ in the region where $\Delta E_{\text{Z}} \gtrsim J$, where $J$ is the energy splitting between $|T_{0}(1,1)\rangle$ and $|S(1,1)\rangle$.

The other method to measure $\Delta E_{\text{Z}}$ is to use singlet-triplet oscillations, where the spins are left to oscillate between the $|S(1,1)\rangle$ state and the $|T_{0}(1,1)\rangle$ states with frequency $\Delta E_{\text{Z}}$. To enable this oscillation, the state needs to be initialised in a $|S(1,1)\rangle$ state. Referring back to Fig. \ref{fig:results}(c), we now ramp quickly through the (0,2)-(1,1) anticrossing, diabatically with respect to the Zeeman splitting differences, but still slowly with comparison to the tunnel rates so that the electrons always end up in the $|S(1,1)\rangle$ configuration. The oscillations initiate immediately after the condition $\Delta E_{\text{Z}} \gg J$ is met. Figure \ref{fig:results}(e) shows the $S-T_{0}$ oscillations as a function of magnetic field angle and the dwell time spent at the (1,1) configuration.

The extracted $\Delta E_{\text{Z}}$ data for both of these methods, shown in Fig. \ref{fig:results}(b), are in excellent agreement. Comparing the rate of blockade lifting and the Zeeman energies in Figs. \ref{fig:results}(a) and (b), it is obvious that $\Gamma_{\text{blockade}}$ increases as $\Delta E_{\text{Z}}$ increases. When $\Delta E_{\text{Z}}$ reaches a minimum at a magnetic field angle of -20 and 100 degrees, there is a minimum in $\Gamma_{\text{blockade}}$ at -30 and 105 degrees. The small deviation is possibly due to Stark shift - $\Delta E_{\text{Z}}$ was measured deep in the (1,1) regime, while $\Gamma_{\text{blockade}}$ is determined at the readout point, where (0,2) is the charge ground state.

\section{Non-unitary processes, predicting blockade rate}
The numerical model is based on the Lindbladian form of the master equation \cite{lindblad_generators_1976,gorini_completely_1976} to allow for the implementation of Markovian noise, details can be found in the Appendix. In this model, an operator is chosen to model the impact of the environment on the quantum system – either a dephasing channel which is a consequence of charge noise, or a phonon-mediated relaxation channel. 

The charge dephasing noise model describes fluctuations in the detuning $\varepsilon$ due to charge fluctuations and how it impacts the energy separation between (1,1) and (0,2) states. In our model, noise in the tunnel rate is not considered because we expect the detuning noise to be the most significant given that it is caused by the dipole component of any far away fluctuators, while the tunnel rate noise is caused by the quadrupole moment~\cite{friesen_decoherence-free_2017}. That said, the quadrupole stray fields may couple exponentially to the tunnel rate, so it is hard to determine without further experiments what would be the leading cause of charge dephasing. We do expect that our results are qualitatively preserved even in the presence of tunnel rate noise, but the Lindbladian model might differ in that case. Furthermore, the dependence of the $T_2$ parameter on the target detuning $\varepsilon$ is not included, which is a good approximation for $\varepsilon\gg t$. Dephasing between spin states in the (1,1) configuration due to fluctuations of $\Delta E_{\text{Z}}$ occur at the order of hundreds of kHz. This is much slower than the scale of noise we are studying here. For the relaxation process, some dependence of $T_1$ on detuning is also expected~\cite{wang_charge_2013}, but we do not include that information in our analysis.

We specifically study the time evolution of the density operator when the initial state is a pure $|T_0\rangle$ state $\hat{\rho}(t=0)=|T_0\rangle\langle T_0|$. A few examples are shown in Fig. \ref{fig:variables}(a) for the case of a dephasing channel with charge coherence time $T_{2}^{\text{charge}}$. It is clear from these plots that there is an exponential damping of the $T_{0}-S$ oscillations, which is marked by red dashed lines showing a fitting curve of $A e^{-t \Gamma_{\text{blockade}}}+B$, which is used to extract $\Gamma_{\text{blockade}}$ from simulations. The evolution of the $|T_0\rangle$ state shows the oscillations between itself and $|S(1,1)\rangle$ becoming damped over time. The dephasing term damps the oscillation until it has reached a fully mixed state. In principle, the model would also describe the overdamped single shot tunnelling across the double dot. In the example given in Fig. \ref{fig:variables}(a) the $|T_0\rangle$ probability saturates at a value of $1/3$. This is because the dephasing channel mixes the three states at equal weights. This is an artifact of a model that only considers dephasing. When relaxation is taken into account, eventually the state is fully polarized in the $|S(0,2)\rangle$ state. Due to the use of a simplified Markovian model, the two effects were not captured together. Instead, the relaxation channel is investigated independently to the dephasing, where $\Gamma_{\text{blockade}}$ values are extracted in a similar way to that described using dephasing channels.

Since neither the charge relaxation or the charge dephasing times are known \textit{a priori} for this device, we study the blockade rate as a function of both parameters in the plots shown in Fig. \ref{fig:variables}(b). The value of the tunnel coupling at this operation configuration is determined to be $t \approx 3$ GHz using a spin-funnel experiment~\cite{petta_coherent_2005}. The detuning of $\varepsilon \approx 140$ GHz is extracted from the operation point and known lever arm (measured using magnetospectroscopy). The Zeeman energy difference varies between $\Delta E_{\text{Z}} \approx 0-20$ MHz (see also Fig. \ref{fig:results}(b)).

\section{Charge relaxation and dephasing times} \label{sec:times}

\begin{figure*}[t]
\includegraphics[width=18cm]{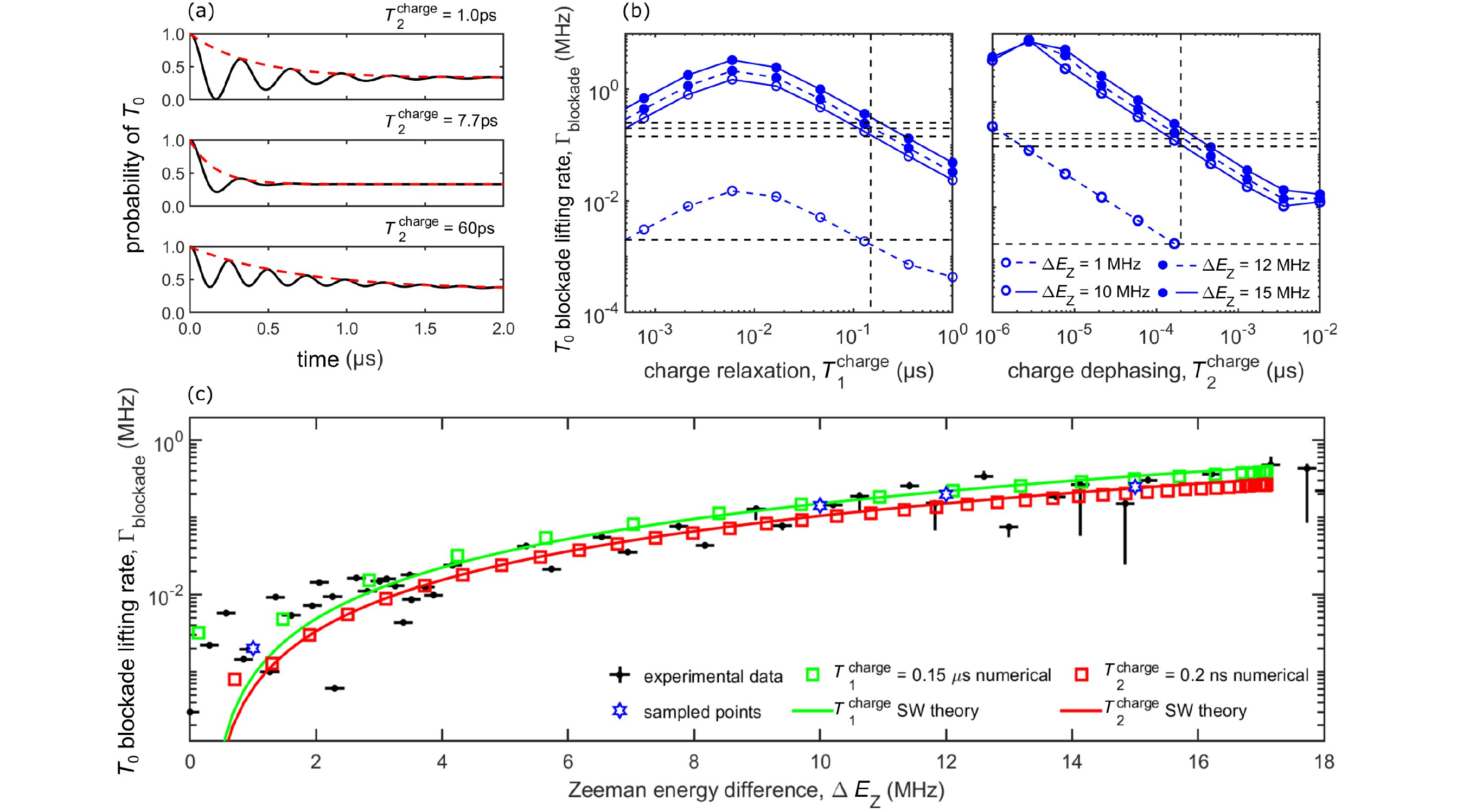}
\centering
\caption{Numerical model for the blockade lifting. (a) For an initial state $|\psi(t=0)\rangle = |T_0 (1,1)\rangle$, the probability as a function of time $p(t)=|\langle T_0 (1,1)|\psi(t)\rangle|^2$ is calculated numerically using Eqs. (\ref{eq:hamiltonian}), (\ref{eq:schrodinger_equation_lind}) and (\ref{eq:dephasing-channel}) including dephasing processes only. The blockade rate $\Gamma_{\text{blockade}}$ can be extracted from the decay (red dashed line). Comparison between all three plots shows that the decay rate is not monotonically dependent on the dephasing time $T_2^{\rm charge}$. (b) The extracted decay rates $\Gamma_{\text{blockade}}$ as a function of charge relaxation and dephasing calculated from the simulation for four values of $\Delta E_{\text{Z}}$ (blue). The values of $\Gamma_{\text{blockade}}$ that correspond to $\Delta E_{\text{Z}}$ from the experimental data (black dashed lines) are then used to extract the relaxation and dephasing times. These are $T_{2}^{\text{charge}}=$ 0.2 ns and $T_{1}^{\text{charge}}=$ 0.15 $\mu$s. For small values of $T_{2}^{\text{charge}}$ and $T_{1}^{\text{charge}}$, the blockade lifting is halted~\cite{burgarth2019generalized}. (c) The comparison of $\Gamma_{\text{blockade}}$ and $\Delta E_{\text{Z}}$ for the experiment (black dots) against the numerical (squares) and analytical (solid line) theoretical decay processes adopting $T_{2}^{\textrm{charge}}=0.2$ ns (red) or $T_{1}^{\textrm{charge}}=0.15$ $\mu$s (green). The blue stars indicate the sampled points used to match the relaxation and dephasing times in (b).}
\label{fig:variables}
\end{figure*}

Experimentally, there is a corresponding $\Gamma_{\text{blockade}}$ value for each $\Delta E_{\text{Z}}$. (The horizontal black dashed lines in Fig. \ref{fig:variables}(b) indicate these). We analyze four values of $\Delta E_{\text{Z}}$ and how they lead to blockade lifting for different values of $T_1$ or $T_2$. Most of these $\Delta E_{\text{Z}}$ values are chosen to be large in order to minimise the impact caused by the Stark shift between the points where $\Gamma_{\text{blockade}}$ and $\Delta E_{\text{Z}}$ are measured. These are used to find the most suited $T_{1}^{\text{charge}}$ and $T_{2}^{\text{charge}}$ values that fit the data accurately (the vertical black dashed lines in Fig. \ref{fig:variables}(b)). From the simulations $T_{2}^{\text{charge}} = 0.2$ ns and $T_{1}^{\text{charge}}=$ 0.15 $\mu$s are found to most accurately fit the data. 

We note that when the charge relaxation or dephasing become very fast, the rate of blockade lifting $\Gamma_{\text{blockade}}$ starts to decrease. The root of this non-monotonic behavior is interpreted as a decoherence-based freezing of the quantum states akin to the quantum Zeno effect~\cite{burgarth2019generalized}, where the Hamiltonian dynamics are halted. We have not considered these smaller values of $T_1$ and $T_2$ since they are not physical, and this regime is not achieved in the experiments presented here.

Values for charge dephasing have been found experimentally to range from 0.127 to 0.760 ns for a charge qubit in a Si/SiGe heterostructure~\cite{shi_coherent_2013}. This is in agreement with $T_{2}^{\mathrm{charge}}$ found in the present work. The relaxation time of a charge qubit in another Si/SiGe device was found to vary depending on the detuning and tunnel rates of the system as well as the geometry of the double quantum dot~\cite{wang_charge_2013}. Assuming the same analysis extends to MOS devices, the expected $T_{1}^{\mathrm{charge}}$ values should be on the order of 10 ms, which would be four orders of magnitude larger than the value that fits our experimental data. 
One cannot completely rule out the relatively short $T_{1}^{\mathrm{charge}}$ concluded here without a direct measurement.  Several nonidealities could impact the real relaxation time in our device, such as oxide imperfections, piezoelectric phonons in silicon dioxide and the complications deriving from the additional electrons in our dots. To confirm this, a direct measurement of $T_{1}^{\mathrm{charge}}$ and $T_{2}^{\mathrm{charge}}$ should be done for a similar device to the one in the present work. This is not possible with our current experimental setup which relies on sub-gigahertz filtering of electric noise.

For the $T_{1}^{\text{charge}}$ and $T_{2}^{\text{charge}}$ values found, the rate of blockade lifting $\Gamma_{\text{blockade}}$ is calculated as a function of the Zeeman energy difference $\Delta E_{\text{Z}}$. This is plotted along side the experimental $\Gamma_{\text{blockade}}$ times in Fig. \ref{fig:variables}(c). The dephasing and relaxation processes both explain the non-linear relationship between $\Gamma_{\text{blockade}}$ and $\Delta E_{\text{Z}}$. In contrast, only relaxation processes matter in GaAs dots \cite{barthel_relaxation_2012}. This is because the singlet states are now coupled not only by the tunnel coupling, but also through the charge oscillation decay mechanism.

\section{Analytical Method}

The quantitative agreement between the model and the experimental results confirms that our interpretation is now complete. We now steer away from numerical simulations and treat the problem analytically. The analytical method gives insight into why the decay rate depends so strongly on the Zeeman energy difference. We can now try to leverage this understanding to design readout schemes that maximize the fidelity of either parity or singlet-triplet readout, depending on the application intended. 

The Schrieffer-Wolff (SW) perturbation theory~\cite{schrieffer_relation_1966} is chosen to study the rate at which the $|T_{0}(1,1)\rangle$ blockade is lifted, the derivation is found in the Appendix. From this, two analytical equations are found to describe $\Gamma_{\text{blockade}}$ as a function of the system parameters. When $\Delta E_{\text{Z}}$ is small the rate becomes, for dephasing effects 

\begin{equation}
    \Gamma_{\text{blockade}} \approx \frac{2\Delta E_{\text{Z}}^{2}}{T_{2}^{\text{charge}}} \frac{t^{2} + \varepsilon^{2}}{t^{2} \varepsilon^{2}},
    \label{eq:schrieffer-wolff-2}    
\end{equation}

and for relaxation effects,

\begin{equation}
    \Gamma_{\text{blockade}} \approx \frac{2\Delta E_{\text{Z}}^{2}}{T_{1}^{\text{charge}}} \frac{t^{2} + \varepsilon^{2}}{t^{4}}.
    \label{eq:schrieffer-wolff-relax}
\end{equation}

It should be noted that $T_{2}^{\text{charge}}$ is dependent on $\varepsilon$ when $\varepsilon$ is comparable or smaller than $t$. In experiments $T_{2}^{\text{charge}}$ is not controllable, but by adjusting the magnetic field of the system $\Delta E_{\text{Z}}$ can be changed. This can be limited, however, since spin-orbit effects might impact other aspects of the qubit control. In that case, one would focus on the electrically controllable $t$ and $\varepsilon$.

\begin{figure}[ht]
    \includegraphics[width=8.6cm]{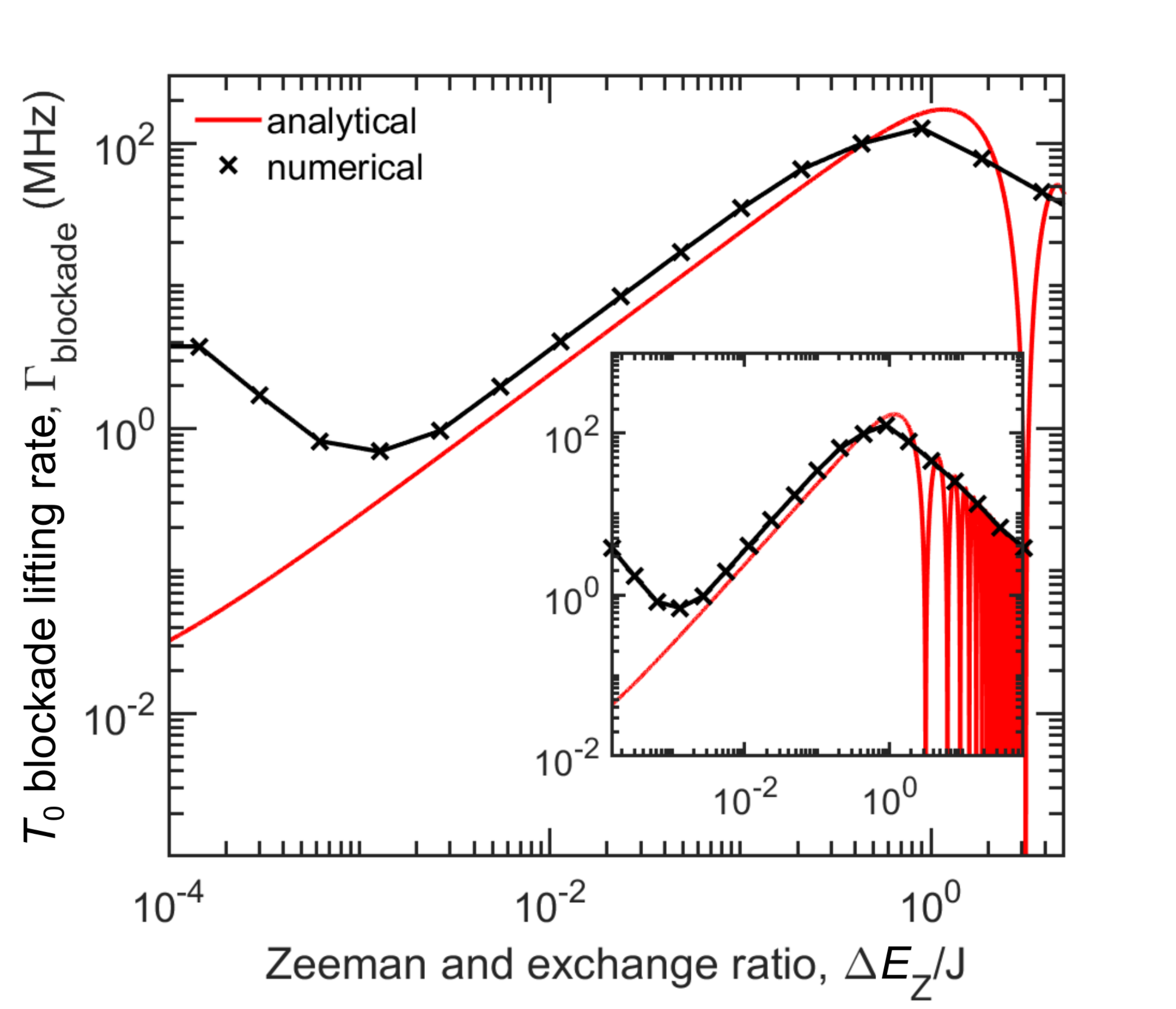}
    \centering
    \caption{Validity limits of the analytical perturbative expression. Comparing $\Gamma_{\text{blockade}}$ obtained by the analytical expression in Eq.~(\ref{eq:schrieffer-wolff}) to the numerical calculations for different values of $\Delta E_{\text{Z}}/J$, where the inset shows this on a larger range. The analytical approximation is shown to be valid for ratios $\Delta E_{\text{Z}}/J$ between $10^{-2}$ and 2. The upper limit is inherent to the perturbative nature of the analysis, while the lower limit relates to the regime where the non-unitary evolution can no longer be treated by the conventional SW method and must be incorporated explicitly in the analysis.}
    \label{fig:J_deltag}
\end{figure}

Comparison between the numerical simulation results for $\Gamma_{\text{blockade}}$ and the analytical expression (\ref{eq:schrieffer-wolff-2}) is shown in Fig. \ref{fig:J_deltag}. In the regime where $\Delta E_{\text{Z}}\gg J$ the SW analysis fails due to the assumption of small splitting between the singlet and triplet sub-spaces needed for the approximation, as mentioned above. Failure of the approximation is also seen at $\Delta E_{\text{Z}}\ll J$ where the non-unitary nature of the system starts to become dominant. This means that the SW method should be adjusted to incorporate the non-unitary evolution~\cite{ficheux2018dynamics}.
 
\section{Discussion}

At first sight, the parity readout scheme may seem to be unable to provide the same level of information as the individual measurements of each spin qubit. This concern is already present with traditional singlet-triplet blockade readout -- the outcome of the readout is a single bit of information (either the electron tunneled through or it did not), while the input was two qubits. 

Measurements on single qubits based on tunnelling to a large reservoir~\cite{elzerman2004single}, however, may be impractical for a dense two dimensional array of qubits in a large scale quantum computer. This means that the latched readout scheme that we used as a tool to freeze the spin dynamics in our work cannot be used in a scalable architecture. This highlights the significance of parity or singlet-triplet readout, meaning the physics of singlet-triplet blockade and the onset of parity readout need to be well understood and controllably reproduced. Some previous literature was dedicated to recovering the most useful protocols for quantum computing using exclusively singlet-triplet readout~\cite{rohling2013tomography}. We describe next two examples that can be adapted for the case of a parity readout; namely, the two qubit state tomography and the syndrome measurements in a quantum error correction protocol.

\subsection{Two qubit state tomography with parity readout} \label{sec:tomography}

State tomography relies on preparing a large ensemble (usually a time ensemble) of nominally identical qubit states and measuring them in different basis sets. In practice, direct measurements in different basis sets is impractical, so the measurement axis is kept fixed and the qubits are manipulated to map the different elements of their density matrices into probabilities of measuring certain outcomes in the fixed basis.

In this sense, universal control of these qubits is always necessary for performing quantum state tomography. The fact that a parity measurement only provides one bit of information as an output is not a significant limitation -- it only means that a larger ensemble needs to be measured in order to obtain all the information in the density matrices. The practical steps to do so using singlet-triplet readout were discussed in Ref.~\onlinecite{rohling2013tomography}. 

The projection operator associated with an odd outcome of the parity readout is $\Pi_{ZZ}^{\rm odd}=(\hat{I}\otimes\hat{I}-\hat{\sigma_Z}\otimes\hat{\sigma_Z})/2$, where $\hat{I}$ is the $2\times2$ identity matrix and the $\hat{\sigma_i}$ denote the Pauli matrices. In order to completely reconstruct a two qubit density operator, the state tomography requires 15 different measurement projections. Given the quantum gates that we have for our qubits, the natural choice for the projections are $\Pi_{MN}=(\hat{I}\otimes\hat{I}-\hat{\sigma_M}\otimes\hat{\sigma_N})/2$, where $\Pi_{MN}$ are linearly independent projection matrices with indices $M, N\in \{I, X, Y, Z\}$. The cases when neither $M$ or $N$ are the identity can be trivially obtained by single qubit rotations, projecting the $X$ or $Y$ components of the spin into the the $Z$ quantisation axis before measuring the parity. In the cases where we want to measure a single qubit, which means that either $M$ or $N$ are the identity $\hat{I}$, it is necessary to perform a CNOT operation between the two qubits before the parity readout.

The estimated density matrix $\hat{\rho}$ can then be reconstructed using 
\begin{equation}
  p_{MN} = Tr[\Pi_{MN}\hat{\rho}],
  \label{parity_readout}
\end{equation}
where $p_{MN}$ are the measurement outcomes (probabilities). 

\subsection{Quantum Error Detection} 

A universal quantum computer with error correction will require a large number of physical qubits and highly accurate measurements of these qubits in order to identify possible errors in the computation. Silicon spin qubits in CMOS devices are small enough that they may be scaled up using the mass production techniques inherited from the transistor industry, but a scalable strategy for reading out each of the physical single spin qubits is challenging. Instead, two spins in a double quantum dot can be used as an ancilla system to detect errors in a logical qubit. 

This idea has been studied in the context of singlet-triplet readout, showing that the surface code implementation can be recovered in a reasonably direct way \cite{jones_logical_2018}. Demonstrating the extension of this analysis to the case of parity readout is trivial. Instead, here we focus on the time scale of the measurements.

The readout time must be much faster than the time it takes for an error to occur. A conservative bound for this time would be the spin coherence time in an echo experiment, which is typically of the order of tens or hundreds of ${\rm \mu}$s. In that case, for a measurement setup similar to ours, one would be able to achieve high fidelity singlet-triplet readout. But, as shown by Eq.~(\ref{eq:schrieffer-wolff-2}), a larger Zeeman splitting difference could result in a measurement that is transitioning between the singlet-triplet regime and the parity regime. In this case, the fidelity of the readout would be compromised. A perhaps counter-intuitive conclusion of our analysis is that by reducing the tunnel rate one might be able to speed up the $|T_0\rangle$ blockade lifting so that the readout time falls comfortably within the parity readout range and the fidelity is improved. In order to understand the necessary range of tunnel rate control necessary to compensate for typical differences in the Zeeman energy, one can compare the systems in Refs.~\onlinecite{Hwang_impact_2017} and  \onlinecite{veldhorst_two-qubit_2015}, which differ in $\Delta E_Z$ by two orders of magnitude. In the former, $\Delta E_Z = 0.41$ MHz \cite{Hwang_impact_2017} so that the tunnel rate should be below 0.14 GHz to be in the parity readout regime, while for the latter the Zeeman energy difference $\Delta E_Z = 17$ MHz sets the transition to be at a  tunnel rate of 17 GHz. These values were estimated assuming all other parameters to be the same as in the present work (the chosen wait time, 1/$\Gamma_{\text{blockade}}$, is 100 $\mu$s, which is typical for DC SET current measurements) and we adopt equation (\ref{eq:schrieffer-wolff-2}). The tunnel coupling values calculated here correspond to the maximum tunnel coupling the system should have to remain in the parity readout regime. This means that if both systems studied here had a tunnel coupling of $0.14$ GHz, they would both exhibit parity readout. The ability to control tunnel coupling by orders of magnitude has been demonstrated using exchange-gate electrodes in quantum dot systems in a variety of material systems \cite{petta_coherent_2005,Zajac_resonantly_2018,russ_high_2018,leon_bellstate_2020}. It is then possible to lower the tunnel rate of all dot pairs in a large system to guarantee that all readout regimes are in parity mode. Potentially the opposite could also be true, depending on how achievable a large enough tunnel coupling is, given the effectiveness of the exchange-gate electrodes and the interdot distances.

\subsection{Summary}

The device studied here allowed us to combine traditional Pauli blockade and latched spin readout to investigate the physical origins of the parity readout. We describe this process in terms of a model that includes both the effects of Zeeman energy difference, as well as the non-unitary charge evolution under the environment-induced noise. We investigated the model numerically and also analytically, using first order perturbation theory, to establish an analytical formula connecting the $|T_0\rangle$ blockade rate with the detuning at the readout point $\varepsilon$, the interdot tunnel rate $t$, the difference in Zeeman splittings $\Delta E_Z$, and either the charge dephasing time $T_{2}^{\text{charge}}$ or the charge relaxation time $T_{1}^{\text{charge}}$. According to our conclusions, high precision two qubit state tomography is viable with this readout scheme. We also showed the pathway for engineering the blockade rate for high fidelity syndrome estimation in a quantum error correction code, revealing that control over the tunnel rate and detuning at the readout point can compensate the Zeeman energy difference.

\section*{Acknowledgments}

We acknowledge support from the US Army Research Office (W911NF-17-1-0198), the Australian Research Council (FL190100167 and CE170100012), Silicon Quantum Computing Pty Ltd, and the NSW Node of the Australian National Fabrication Facility. The views and conclusions contained in this document are those of the authors and should not be interpreted as representing the official policies, either expressed or implied, of the Army Research Office or the U.S. Government. The U.S. Government is authorized to reproduce and distribute reprints for Government purposes notwithstanding any copyright notation herein. B. H. acknowledges support from the Netherlands Organization for Scientific Research (NWO) through a Rubicon Grant. K. M. I. acknowledges support from a Grant-in-Aid for Scientific Research by MEXT. K. Y. T. acknowledges the support from the Academy of Finland (grant numbers 308161, 314302 and 316551). The authors declare that they have no competing financial interests.

\section*{Appendix: Methods}

\subsection{Numerical model}
The model Hamiltonian includes the tunnel coupling $t$ between the singlet states and the mixing of the $|T_{0}(1,1)\rangle$ and $|S(0,2)\rangle$ through the $|S(1,1)\rangle$ state, and the coupling between the $|S(1,1)\rangle$ and $|T_{0}(1,1)\rangle$ states because of the difference of Zeeman energies between the dots. In the basis $\{|S(0,2)\rangle, |S(1,1)\rangle, |T_{0}(1,1)\rangle\}$ our Hamiltonian is
\begin{equation}
\hat{H} = 
    \begin{pmatrix} 
    -\varepsilon & t & 0 \\
    t & 0 & \Delta E_{\text{Z}} \\
    0 & \Delta E_{\text{Z}} & 0
    \end{pmatrix},
    \label{eq:hamiltonian}
\end{equation}
where $\varepsilon$ is the detuning between the dots. 

We study the impact of noise on the system by simulating the time evolution of the density matrix $\hat{\rho}$ as a master equation in the Lindbladian form~
\cite{lindblad_generators_1976,gorini_completely_1976}
\begin{equation}
    \frac{d \hat{\rho}}{dt} = -i[\hat{H},\hat{\rho}] + \hat{\hat{\mathcal{L}}}[\hat{a}](\hat{\rho}).
    \label{eq:schrodinger_equation_lind}
\end{equation}
The Lindblad superoperator $\hat{\hat{\mathcal{L}}}[\hat{a}](\hat{\rho})$ acts on $\hat{\rho}$, describing the non-unitary evolution of the open quantum system under an assumed Markovian noise. The operator $\hat{a}$ is a jump operator chosen to model the impact of the environment on the quantum system -- either a dephasing channel which is a consequence of charge noise, or a phonon-mediated relaxation channel. The Lindblad superoperator part of the master equation can be expanded as 
\begin{equation}
    \hat{\hat{\mathcal{L}}}[\hat{a}](\hat{\rho}) = \hat{a} \hat{\rho} \hat{a}^{\dagger} - \frac{1}{2}(\hat{a} \hat{a}^{\dagger} \hat{\rho} + \hat{\rho} \hat{a} \hat{a}^{\dagger}). 
    \label{eq:master}
\end{equation}
Using frequency units, Eq. (\ref{eq:master}) can be used to describe charge dephasing, $\hat{a}_{\text{dephasing}}$, and charge relaxation processes, $\hat{a}_{\text{relaxation}}$, using
\begin{equation}
\hat{a}_{\text{dephasing}} = 
    \begin{pmatrix} 
    \frac{1}{\sqrt{2T_{2}}} & 0 & 0 \\
    0 & -\frac{1}{\sqrt{2T_{2}}} & 0 \\
    0 & 0 & -\frac{1}{\sqrt{2T_{2}}}
    \end{pmatrix},
    \label{eq:dephasing-channel}
\end{equation}
or
\begin{equation}
\hat{a}_{\text{relaxation}} = 
    \begin{pmatrix} 
    0 & \frac{1}{\sqrt{T_{1}}} & 0 \\
    0 & 0 & 0 \\
    0 & 0 & 0
    \end{pmatrix}.
    \label{eq:relaxation-channel}
\end{equation}
The operator $\hat{a}_{\text{dephasing}}$ models the phase noise acting on the charge states (1,1) and (0,2) charge states. The operator $\hat{a}_{\text{relaxation}}$ models the depopulation of $|S(1,1)\rangle$ relaxing into the $|S(0,2)\rangle$ state.

\subsection{Analytical model}
We adopt quasi-degenerate perturbation theory~\cite{schrieffer_relation_1966} to study the dynamics of the $|T_0\rangle$ unblocking as a leakage into the singlet sector of the total three dimensional Hilbert space. For this approach to be valid, all off-diagonal terms in the Hamiltonian must be small compared to the splitting between the triplet and the singlet sectors. To ensure this condition is satisfied even when $t$ is comparable to $\varepsilon$, a unitary transformation is used to bring the Hamiltonian to a basis with symmetric and anti-symmetric singlet state combinations.  To achieve this partial diagonalisation, the unitary transformation $\hat{U}^{\dagger}\hat{H_{0}}\hat{U}$ is performed, with a suitable choice of $\hat{U}$
\begin{equation}
    \hat{U} = 
    \begin{pmatrix}
    \frac{t\Delta_{S-T_{0}}^{\mathrm{antisym}}}{\Delta E_{\text{Z}}( \varepsilon + \sqrt{4t^{2} + \varepsilon^{2}} )} & \frac{\Delta_{S-T_{0}}^{\mathrm{antisym}}}{\Delta E_{\text{Z}}} & 0 \\
    \frac{1}{\sqrt{2 (1+\varepsilon(\sqrt{4t^{2}+\varepsilon^{2}})^{-1})^{-1}}} & \frac{\Delta_{S-T_{0}}^{\mathrm{sym}}}{\Delta E_{\text{Z}}} & 0 \\
    0 & 0 & 1
    \end{pmatrix},
\end{equation}
to give
\begin{equation}
    \hat{H}' =
    \begin{pmatrix} 
    \frac{-\varepsilon}{2}+\sqrt{\frac{\varepsilon^{2}}{4}+t^{2}} & 0 & \Delta_{S-T_{0}}^{\mathrm{antisym}} \\
    0 & \frac{-\varepsilon}{2}-\sqrt{\frac{\varepsilon^{2}}{4}+t^{2}} & \Delta_{S-T_{0}}^{\mathrm{sym}} \\
    \Delta_{S-T_{0}}^{\mathrm{antisym}} & \Delta_{S-T_{0}}^{\mathrm{sym}} & 0
    \end{pmatrix}
    \label{eq:hamiltonian_diag}
\end{equation}
\noindent where
\begin{equation}
       \Delta_{S-T_{0}}^{\mathrm{antisym}} = \frac{\Delta E_{\text{Z}} \big( \varepsilon + \sqrt{4t^{2} + \varepsilon^{2}}\big )}{\sqrt{8t^{2} + 2\varepsilon \big(\varepsilon + \sqrt{4t^{2} + \varepsilon^{2}} \big )}},
\end{equation}
and
\begin{equation}
     \Delta_{S-T_{0}}^{\mathrm{sym}} = \frac{\Delta E_{\text{Z}} \big( \varepsilon - \sqrt{4t^{2} + \varepsilon^{2}}\big )}{\sqrt{8t^{2} + 2\varepsilon \big(\varepsilon - \sqrt{4t^{2} + \varepsilon^{2}} \big )}}.
\end{equation}

Splitting the Hamiltonian into the sum
\begin{equation}
    \hat{H} = \hat{H_{0}} + \hat{H_{1}} + \hat{H_{2}},
\end{equation}

\noindent one can now determine the unitary operator $e^{\hat{S}}$ that approximately diagonalises the Hamiltonian to first order in the small perturbations $\Delta_{S-T_{0}}^{\mathrm{antisym}}$ and $\Delta_{S-T_{0}}^{\mathrm{sym}}$ following the usual SW algorithm. 

We are more interested in the damping of the $|T_0\rangle$ population introduced by the Lindblad superoperator. We use the fact that the Lindbladian equation is invariant under unitary transformations and obtain the transformed quantum channel
\begin{equation}
    \hat{a'} = e^{-\hat{S}} \hat{U}^{\dagger} \hat{a} \hat{U} e^{\hat{S}}.
    \label{eq:transform_a}
\end{equation}
This enables us to find the damping of $|T_0\rangle$ as a function of the system parameters by looking at the resulting dynamical equation for $\rho_{T_{0}} = \langle T_{0}(1,1)| \hat{\rho} |T_{0}(1,1)\rangle$. The general form for this equation is
\begin{equation}
    \frac{d \rho_{T_{0}}}{dt} = -\Gamma_{\text{blockade}}\rho_{T_{0}}+C,
    \label{eq:rhoT0}
\end{equation}
where $C$ stands for the terms that are not proportional to $-\rho_{T_{0}}$, and therefore are not responsible for damping.

Substituting Eqs. (\ref{eq:transform_a}) and (\ref{eq:rhoT0}), for the dephasing jump operator (\ref{eq:dephasing-channel}), into the master equation (\ref{eq:master}) and looking at the $\rho_{T_{0}}$ elements only, the analytical expression is found,
\begin{equation}
    \Gamma_{\text{blockade}} \approx \frac{2t^{2}}{T_{2}^{\text{charge}} \varepsilon^{2}} \sin \bigg( \frac{\Delta E_{\text{Z}} \sqrt{t^2+\varepsilon^{2}}}{t^{2}} \bigg) ^{2}.
    \label{eq:schrieffer-wolff}
\end{equation}
This expression is useful because it gives us insight into how to control $\Gamma_{\text{blockade}}$. Specifically, when $\Delta E_{\text{Z}}$ is small, Eq. (\ref{eq:schrieffer-wolff}) becomes
\begin{equation}
    \Gamma_{\text{blockade}} \approx \frac{2\Delta E_{\text{Z}}^{2}}{T_{2}^{\text{charge}}} \frac{t^{2} + \varepsilon^{2}}{t^{2} \varepsilon^{2}}.
\end{equation}

For the case of the relaxation process, $\Gamma_{\text{blockade}}$ is found using the same SW method and, when $\Delta E_{\text{Z}}$ is small, one obtains

\begin{equation}
    \Gamma_{\text{blockade}} \approx \frac{2\Delta E_{\text{Z}}^{2}}{T_{1}^{\text{charge}}} \frac{t^{2} + \varepsilon^{2}}{t^{4}}.
\end{equation}

%\section*{Author contributions}

%A.E.S. performed the theoretical simulations and analytical methods under the supervision of A.S. and input from S.N.C.. R.C.C.L. and A.S. discuss the impact on quantum computation. T.T. performed the experiments, with R.Z. and B.H. contributing to the preparation of experiments. K.Y.T. and F.H. fabricated the sample. J.Y., C.H.Y., A.M., A.L., and S.N.C. contributing to results discussion and interpretation. A.E.S, T.T., A.S. and A.S.D. wrote the paper with input from all co-authors.

\bibliography{blockade}

\end{document}